\documentclass{WileyMSP-template}
\usepackage{textcomp}
\usepackage{siunitx}
\usepackage[T1]{fontenc}
\usepackage{bm}
\usepackage{xcolor}
\usepackage{amsmath,amssymb,amsthm}

\usepackage{ragged2e}
\justifying
\usepackage{hyperref}

\begin{document}

\pagestyle{fancy}
\rhead{\includegraphics[width=2.5cm]{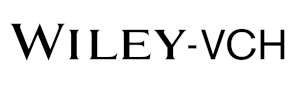}}

\title{Generation of C-band entangled photon pairs by biexciton-exciton cascade from symmetric InAs/InP quantum dots}


\maketitle


\author{Anna Musia{\l}*}
\author{Maja Wasiluk}
\author{Katarzyna Roszak}
\author{Micha{\l} Gawe{\l}czyk}
\author{Wojciech Rudno-Rudzi{\'n}ski}
\author{Pawe{\l} Wyborski}
\author{Johann P. Reithmaier}
\author{Mohamed Benyoucef}
\author{Grzegorz S\k{e}k}


\dedication{}

\begin{affiliations}
A. Musia{\l}, M. Wasiluk, W. Rudno-Rudzi{\'n}ski, Wyborski, G. S\k{e}k\\
Laboratory for Optical Spectroscopy of Nanostructures, Department of Experimental Physics, Wroc{\l}aw University of Science and Technology, Wybrzeże Wyspiańskiego 27, 50-370 Wroc{\l}aw, Poland\\ 
Email Address: anna.musial@pwr.edu.pl

K. Roszak\\
FZU - Institute of Physics of the Czech Academy of Sciences, Na Slovance 1999/2, 182 00 Prague 8, Czechia 

M. Gawe{\l}czyk\\
Institute of Theoretical Physics, Wroc{\l}aw University of Science and Technology, Wybrzeże Wyspiańskiego 27, 50-370 Wroc{\l}aw, Poland

J. P. Reithmaier, M. Benyoucef\\
Institute of Nanostructure Technologies and Analytics (INA), Center for Interdisciplinary Nanostructure Science and Technology (CINSaT), University of Kassel, Heinrich-Plett-Street 40, 34132 Kassel, Germany

\end{affiliations}



\begin{abstract}

Hereby, we study the generation of pairs of polarization-entangled photons at telecom C-band by biexciton-exciton cascade from non-resonantly excited epitaxial InAs/InP quantum dots (QDs). It is realized without external tuning of the fine structure splitting (FSS), which does not exceed 10 ${\mu}$eV in as-grown nanostructures, due to their high in-plane symmetry. Excitonic complexes are identified by means of excitation power-dependent and polarization-resolved (magneto)microphotoluminescence. Their origin from different carrier configurations confined in the same QD is confirmed by time correlations of emitted photons. Experimental results are supported by 8-band $\bm{k}\cdot\bm{p}$ calculations, followed by the configuration interaction method to include excitonic effects. This comparison reveals the structure of higher energy states, allowing for the reconstruction of the QD structural parameters. To verify and quantify the entanglement, we perform quantum state tomography and reconstruct the two-photon density matrix. Its diagonalization allows a detailed analysis of the entangled state and reveals an unfavourable interplay between phase accumulation and decoherence, pointing to a clear route for boosting entanglement by using a XX resonant, pulsed excitation scheme and shortening the radiative lifetimes.

\end{abstract}


\section{Introduction}
Non-classical light states are an important resource for quantum technologies. In particular, single photon sources can enable the realization of secure quantum key distribution (QKD) \cite{BB84,Sangouard2007,Huwer2017} or linear quantum computing \cite{Kiraz2004}. Even more compellingly, photon entanglement opened up the possibility to realize quantum teleportation \cite{Bennett1993}, quantum repeaters \cite{Briegel998}, further QKD protocols \cite{Ekert1991}, all leading to a global quantum network \cite{Kimble2008}. To minimize losses, both in silica fibers and in free-space satellite transmission, photons from the 3rd telecommunication window, centered at 1550~nm, are required. It has been repeatedly proven that these application-relevant quantum light states can be generated by quantum dots (QDs) \cite{Akopian2006, Michler2000, Kurtsiefer2000, Muller2018, Olbrich2017, Zeuner2021, Shields2007}.

GaAs-based QDs, emitting at shorter wavelengths, have already outperformed other types of quantum emitters in terms of the quality of emitted states (degree of entanglement, indistinguishability, or probability of multiphoton events) and source brightness \cite{Huber2018, Zhai2022, Liu2019}. However, there is still much to explore when it comes to QDs emitting at telecom wavelengths, both on a fundamental level and in optimizing photon generation. There are different material platforms considered for telecom emission: strain-engineered GaAs-based \cite{Semenova2008, Joos2024, Nawrath2021, Zeuner2021, Vyvlecka2023}, InAs(P)/InP \cite{Muller2018, Anderson2020, Laccotripes2025, Laccotripes2024, Takemoto2004, Takemoto2007, Holewa2024, Holewa2022, Vajner2024, Kors2018, Carmesin2017, Yacob2014, Barbiero2022}, InP/In$_y$Al$_{1-y}$As/In$_x$Ga$_{1-x}$As \cite{Deutsch2023} and recently also GaSb-based \cite{Chellu2021, Michl2023, Hakkarainen2024}. Interestingly, various growth techniques (metalorganic chemical vapor deposition, molecular beam epitaxy - MBE, chemical beam epitaxy) and growth modes (Stranski-Krastanow, droplet epitaxy, local nanohole droplet etching) are explored to optimize the performance of these quantum emitters. 

In this work, we theoretically describe and experimentally determine the properties of excitonic complexes in InAs(P)/InP QDs exhibiting high in-plane symmetry owing to the additional ripening step during MBE growth. They have already been proven to generate single photons with a low probability of multiphoton events \cite{Musial2020} and extraction efficiency exceeding 13\% when placed in a cylindrical mesa of optimized design \cite{Musial2021, Mrowinski2019}. The emission spectrum of a QD is typically dominated by a negatively charged trion \cite{Podemski2021, Rudno2021, Wasiluk2025}, but here we focus on the neutral biexciton-exciton cascade and its potential to generate pairs of polarization-entangled photons.  


\section{Experimental}
\begin{figure}[tb]
  \includegraphics [width=0.6\linewidth] {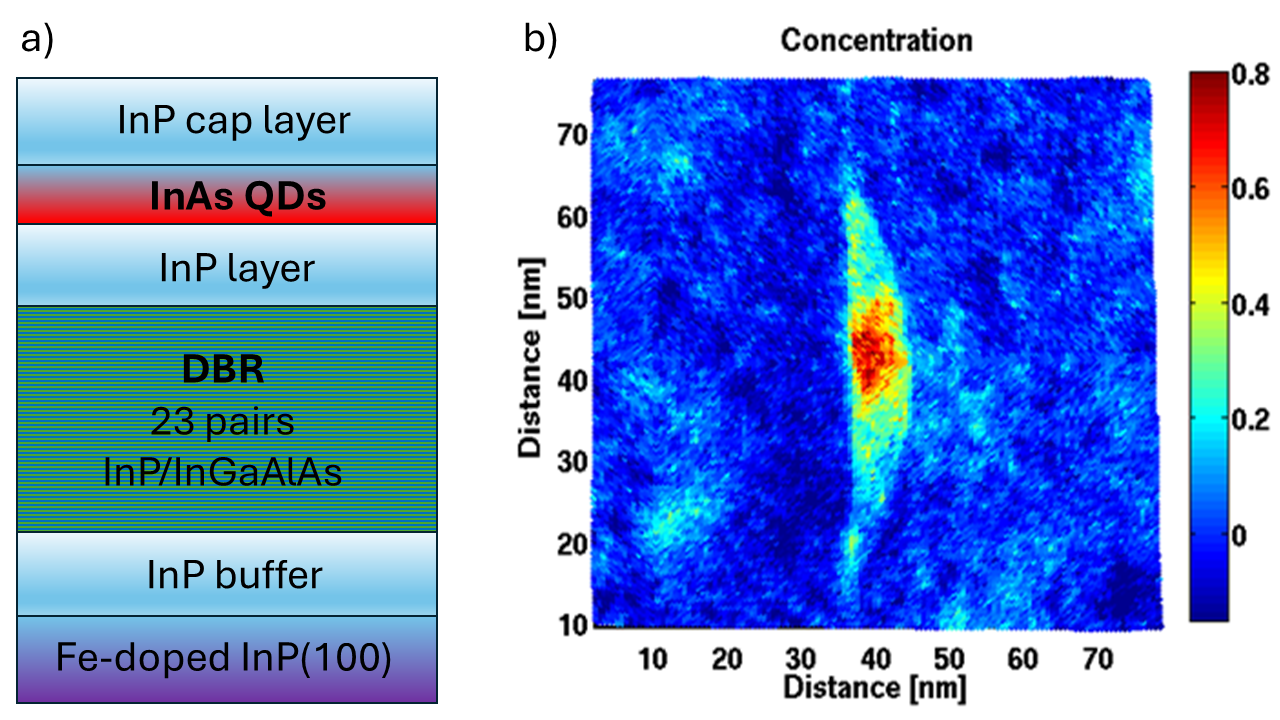}
  \caption{a) Layer design of the investigated InAs/InP quantum dot structure; b) Image obtained via transmission electron microscopy with energy-dispersive X-ray mapping, with arsenic content $x$ color coded.}
  \label{fig:TEM}
\end{figure}

\threesubsection{Investigated structure} The structure under study consists of InAs(P)/InP epitaxial QDs grown by the ripening-assisted MBE technique \cite{Kamins1999} in the Stranski-Krastanow mode directly in an InP matrix \cite{Yacob2014, Kors2018}. This results in a low QD surface density of $\sim10^8$/cm$^2$ and large QDs (with 10-15~nm height and in-plane diameter in the range of 40-70~nm), exhibiting typically high in-plane symmetry. Due to diffusion between the QD material and the substrate, the QDs are, in fact, made of InAs$_x$P$_{1-x}$, with strongly inhomogeneous composition within the QD volume visible in the cross-sectional transmission electron microscopy (TEM) image shown in \textbf{Figure~\ref{fig:TEM}b}. To enhance photon extraction efficiency, a bottom InGaAlAs/InP distributed Bragg reflector (DBR) precedes the growth of the QD layer \cite{Kors2018, Smolka2021}. Cylindrical micrometer-sized mesas etched on the sample surface facilitate investigation of the same nanostructures in various experimental setups and limit the number of excited QDs.\\

\threesubsection{Microphotoluminescence} Excitation power-dependent and polarization-resolved single QD emission is studied in a standard microphotoluminescence ({\textmu}PL) setup featuring: continuous wave (cw) non-resonant excitation with a semiconductor laser at 640~nm (maximal excitation power of 4~mW), a long working distance achromatic infinity-corrected microscope objective with the 0.4 numerical aperture, a liquid helium flow micro-cryostat, a spectrometer composed of a 1~m focal-length monochromator and a liquid N$_2$ cooled multichannel linear InGaAs detector providing spectral resolution of 20~${\mu}$eV. To perform the microphotoluminescence excitation ({\textmu}PLE) experiment, a cw tuneable (1425--1530~nm) external cavity laser, filtered by a 30~cm focal-length monochromator and a set of premium longpass and bandpass filters in the detection path, is used.\\

\threesubsection{Magnetooptical measurements} For magnetooptical experiments, the sample was mounted in an optical cryostat with a superconducting coil magnet providing a magnetic field of up to 5~T in the sample growth direction. To apply the field in the Voigt configuration, the sample is mounted in a perpendicular orientation on a copper block, together with a silver-protected mirror oriented at 45 degrees. The other parameters are as in the {\textmu}PL setup.\\

\threesubsection{Quantum state tomography} A correlation spectroscopy setup is used for cross-correlation measurements proving that certain single QD emission lines in the spectrum originate from radiative recombination of different carrier configurations confined within the same QD, as well as for the quantum state tomography. For quantum state tomography, additional polarization optics is inserted into each arm of the free-space Hanbury Brown and Twiss interferometer, namely quarter-, half-wave plates, and a linear polarizer, to project the polarization of emitted photons onto the two-photon state polarization basis. An additional half-wave plate is used to rotate the polarization of emitted photons onto the laboratory reference frame for polarization measurements. The polarization characteristics of the experimental setup and the polarization optics are determined using a polarimeter. The optical excitation is provided by a cw non-resonant semiconductor laser at 660~nm. It is focused to a spot with a diameter of 1~${\mu}$m on the sample surface by an infinity-corrected microscope objective with 0.4 numerical aperture. The same objective collects the emission and guides it to a non-polarizing 50:50 cube beam splitter, with a 32~cm focal length monochromator on each output. Polarization optics to set the measurement basis is placed in front of each of the two monochromators. After each monochromator, the spectrally-filtered emission (0.45~nm bandwidth, set by the size of the exit slit of the monochromator) is coupled via a single-mode fiber to a superconducting nanowire single photon detector (SNSPD), featuring quantum efficiency exceeding 85\% at 1.55~${\mu}$m, dark counts on the level of 100~Hz and timing jitter better than 50~ps. The operation temperature of the SNSPD is 1.8~K, and polarization paddles at the input fiber are used to maximize the absorption of photons by the NbN active region (polarization parallel to the nanowires). For the quantum state tomography measurements, an excitation power of 1.5~${\mu}$W and a time-bin width of 256~ps is used. No deconvolution of measured data is applied.\\

\section{Excitonic complexes}
\subsection{Experimental results}

\begin{figure}[tb]
  \includegraphics[width=0.8\linewidth]{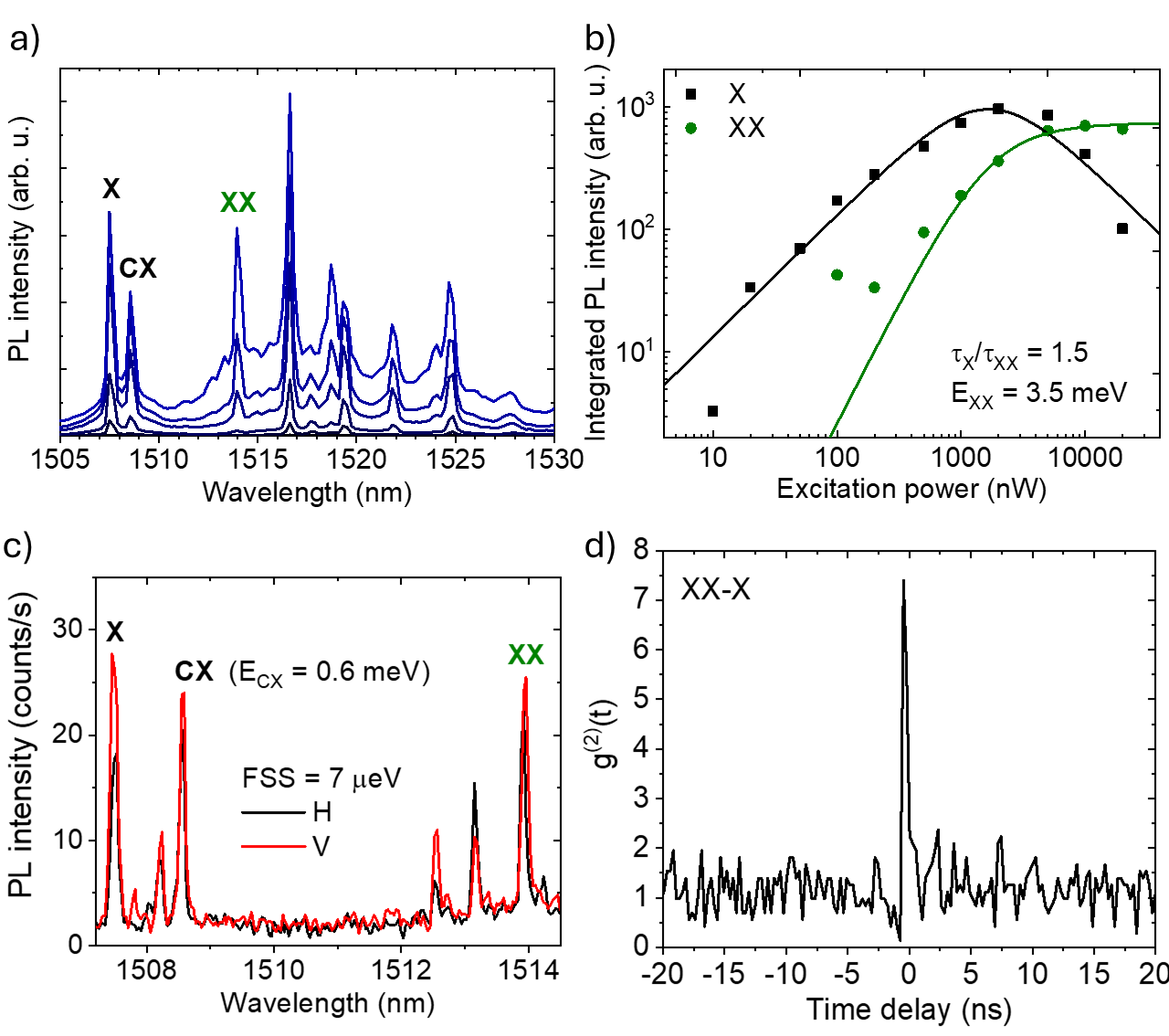}
  \caption{Low temperature (5K) measurements of a single QD: a) Selected {\textmu}PL spectra for various excitation powers from 10~nW to 5~{\textmu}W with basic excitonic complexes, exciton (X), charged exciton (CX), and biexciton (XX), marked; b) Integrated emission intensity (from Gaussian fits to individual emission lines) of X (black squares) and XX (green dots) as a function of excitation power with fit to experimental data (solid lines) with solution of the three-level rate equation model and X to XX lifetime ratio of 1.5; c) {\textmu}PL spectra for two orthogonal linear polarization directions (H - horizontal and V - vertical in the reference frame of the excitonic states) measured for 5~{\textmu}W excitation power; d) normalized cross-correlation histogram of the biexciton-exciton cascade at 5~{\textmu}W excitation power.}
  \label{fig:excitonic}
\end{figure}

To identify basic excitonic complexes, a series of {\textmu}PL measurements as a function of excitation power in the range of 10~nW to 20~{\textmu}W is performed. The exemplary spectra from this series are presented in \textbf{Figure~\ref{fig:excitonic}a}. 

With increasing excitation power, the intensity of emission lines increases, and more emission lines appear in the spectrum, e.g., the one marked as the biexciton (XX), which is clearly visible only from 100~nW excitation power. This is because excitonic complexes consisting of more carriers require higher excitation power (corresponding to a higher density of carriers in the surrounding of the QD available to be captured) to be formed. The power dependence of exciton (X) and XX emission intensities as a function of excitation power exhibits characteristic behavior that we fit with a rate equation model, assuming the steady state occupations of X and XX states and neglecting other possible carrier states confined in the QD as well as spin degree of freedom \cite{Sek2010}. We obtain the best fit to the experimental data for the X to XX lifetime ratio of 1.5, which is lower than the typical value for the strong confinement regime equal to 2 \cite{Wimmer2006, Narvaez2006}. This is consistent with the relatively large volume of the QDs and their dense ladder of excited states \cite{Podemski2021}.  
The lifetime ratio is independently confirmed in a time-correlated single-photon counting experiment under non-resonant excitation (not shown here), where low excitation low-temperature measurements yielded single-exponential decay with lifetimes of $\sim1.2$~ns for X and $\sim0.8$~ns for XX. To confirm that these two emission lines originate from the recombination of neutral excitonic complexes confined in the same QD, a linear-polarization resolved {\textmu}PL \textbf{Figure~\ref{fig:excitonic}c} and cross-correlation data \textbf{Figure~\ref{fig:excitonic}d} are collected. Fitting the emission lines with a Gaussian function for different linear polarization directions allowed us to determine the exciton fine structure splitting (FSS) to be of $\sim7$~{\textmu}eV, in agreement with the high in-plane QD shape symmetry, but hinting at some residual asymmetry of the confining potential
due to the piezoelectric potential
or alloying randomness in the QD. To illustrate the very low FSS, two spectra for orthogonal linear polarization directions are presented in \textbf{Figure~\ref{fig:excitonic}c}. The splitting is determined from a sine fit to the dependence of emission energy on the direction (angle) of detected linear polarization. The X and XX exhibit the same splitting within the experimental accuracy, which, together with the anti-phase behavior (the same polarization of the two inner lines and the two outer ones), indicates their origin from the same QD. Strong bunching visible in the XX-X coincidences histogram additionally proves
cascaded emission from the same QD. All this leads up to the conclusion that we observe X and XX from the same QD and, therefore, allows determining its biexciton binding energy to be 3.5~meV, typical for InAs/InP nanostructures \cite{Sek2009, HolewaPRB2020}. The cross-correlation of pairs of emission lines present in the spectrum (not shown here) revealed one more emission line originating from the same QD marked as CX in the spectra. The cross-correlation between this emission line and the neutral X showed asymmetric anti-bunching, pointing out mutually exclusive excitonic complexes. Combining this with the lack of fine structure splitting, the linear increase in the emission intensity as a function of excitation power, and emission in the low excitation power range suggests that this is emission from a charged exciton. Without a gated structure and controlled charging of the QD, the sign of the excess charge cannot be undoubtedly claimed. The n-doping of the substrate and the presence of the negative circular polarization in this sample \cite{Podemski2021, Wasiluk2025} suggest that it is rather the negatively charged exciton. Further arguments will be brought by the comparison of the experimentally determined binding energy of 0.6~meV with calculations of the excitonic states (see the following Subsection). 

To differentiate between neutral and charged excitons in the case of low FSS, a {\textmu}PL measurements in the external magnetic field in Voigt configuration can be used \cite{Bennett2013, Yamamoto2009}. To comprehensively characterize the QD under study, measurements in two orthogonal directions of the magnetic field up to 5~T are performed (F -- Faraday and V -- Voigt configurations). In this particular case, the splittings for the in-plane magnetic field are too small, relative to the emission linewidth in the range of 150~{\textmu}eV, to resolve them within the spectral resolution of the experimental setup. The diamagnetic shift could be determined only for the neutral exciton, with the value of the diamagnetic coefficient of $5 \times 10^{-7}$~{\textmu}eV/T$^{2}$, corresponding to the characteristic length of 2.5 nm (the effective masses are taken as the same for both directions). In the strong confinement regime, this would be interpreted as the vertical extension of the wave function \cite{Walck1998}. A much smaller value compared to the physical QD height (10~nm) suggests that it is rather determined by an inhomogeneous As distribution. However, if the confinement is weaker, the interpretation shifts toward the distance between the centers of mass of the electron and the hole densities and suggests the separation of the two in the vertical direction \cite{Walck1998}, which, according to calculations, is $\sim0.4$~nm. \textbf{Table~\ref{parameters}} summarizes all the experimentally determined parameters of the three basic excitonic complexes.

\begin{table}
	\caption{\label{parameters} Summary of the experimentally-determined (magneto-)optical properties of excitonic complexes in the investigated single QD. In the second column, emission energy (binding energy) is given for X (CX and XX). The diamagnetic shift and $g$-factor are given for two measurement configurations of the magnetic field: Faraday (out of plane)/Voigt (in-plane)}
	\begin{tabular}[tbp]{@{}|c|c|c|c|c|c|@{}}
		\hline		
		QD&Energy (meV)&FSS ({\textmu}eV)&Lifetime (ns)& Diamagnetic shift ({\textmu}eV/$T^2$)&g-factor\\
		\hline
		X&822.5&7&1.2&1.41/0.05&0.65/0\\
		CX&0.6&&1.1&1.28/0&0.68/0\\
		XX&3.5&&0.8&1.34/0&0.66/0\\
		\hline
	\end{tabular}
\end{table}

\begin{figure}[tb]
  \includegraphics[width=0.9\linewidth]{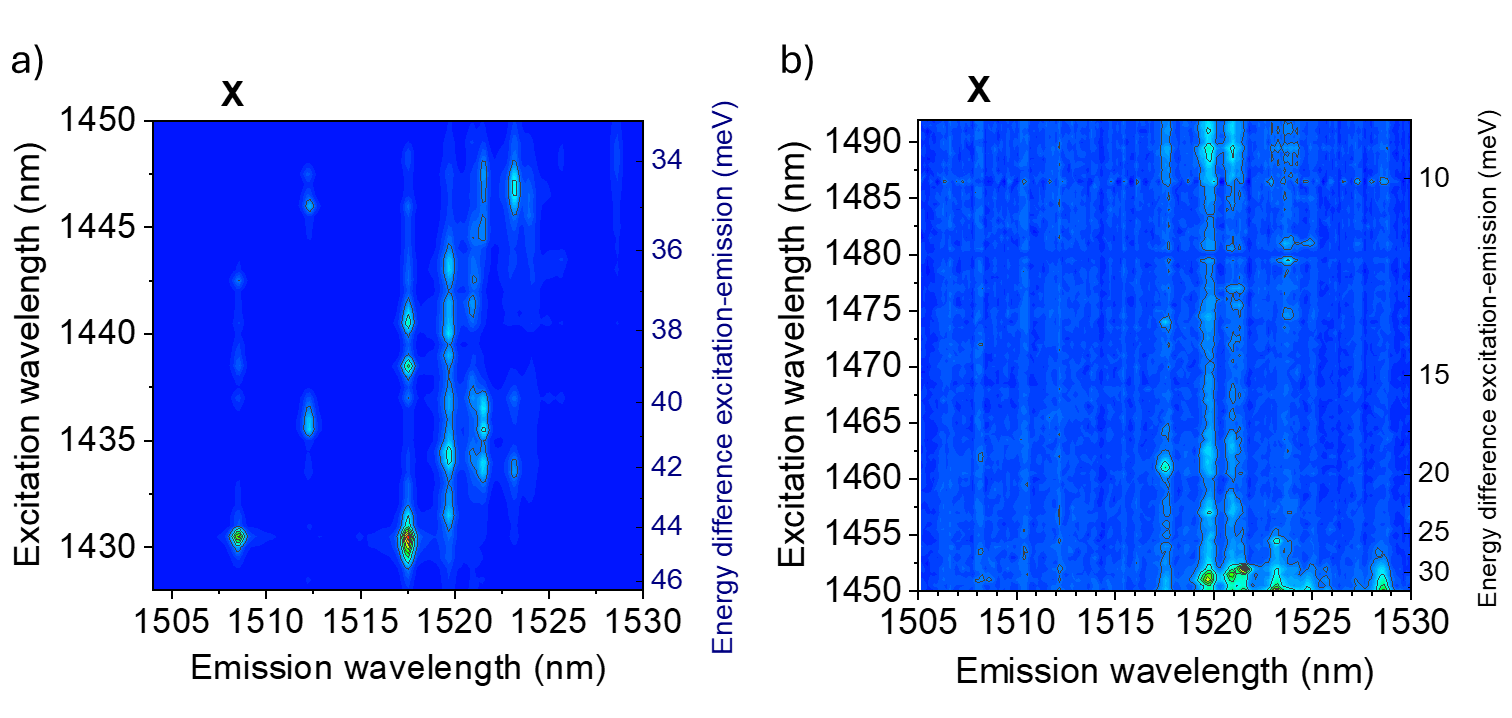}
  \caption{Low temperature (5K) {\textmu}PLE map measured at excitation power of 500~{\textmu}W} for a) short excitation wavelength (high excitation energy); b) long excitation wavelength (low excitation energy) with exciton emission line marked
  \label{fig:uPLE_map}
\end{figure}

To provide more insight into the energy structure of the QD, the energies for efficient QD excitation energies are determined using absorption-like {\textmu}PLE technique with results shown in \textbf{Figure~\ref{fig:uPLE_map}}.  

The excitation wavelength is varied from 1425~nm to 1495~nm, corresponding to the energy range of (0.87--0.83)~eV, with a step of 0.5~nm (approx. 300~{\textmu}eV) and excitation power of 500~{\textmu}W. For the QD under study, only the exciton exhibits a strong resonance at 1430.5~nm (0.867~eV), separated from the emission by 44~meV, which has to be compared with the sum of the calculated electron and hole level separations. This resonance could be related to the phonon-assisted transition, with 42.8~meV (room temperature value) LO-phonon energy in the surrounding InP matrix \cite{Palenskis2014} or to higher energy states. There are three close together lower energy resonances visible in the {\textmu}PLE map at 1436.9~nm (0.863~eV), 1438.6~nm (0.862~eV), and 1442.5~nm (0.860~eV), suggesting a rather dense ladder of excited states. Interestingly, no clear resonances are observed up till the highest experimentally available excitation energy --- no optically active states have been spotted in the range of 30~meV from 0.86~eV to 0.83~eV, as close to the emission energy as 7~meV. The lack of {\textmu}PLE resonances may be due to either the lack of optically active transitions in this spectral range or the phonon bottleneck for exciton relaxation with the given energy difference (the energy is below LO phonon energy and above the range of efficient acoustic phonon-driven processes). 

\begin{figure}[tb]
  \includegraphics[width=0.9\linewidth]{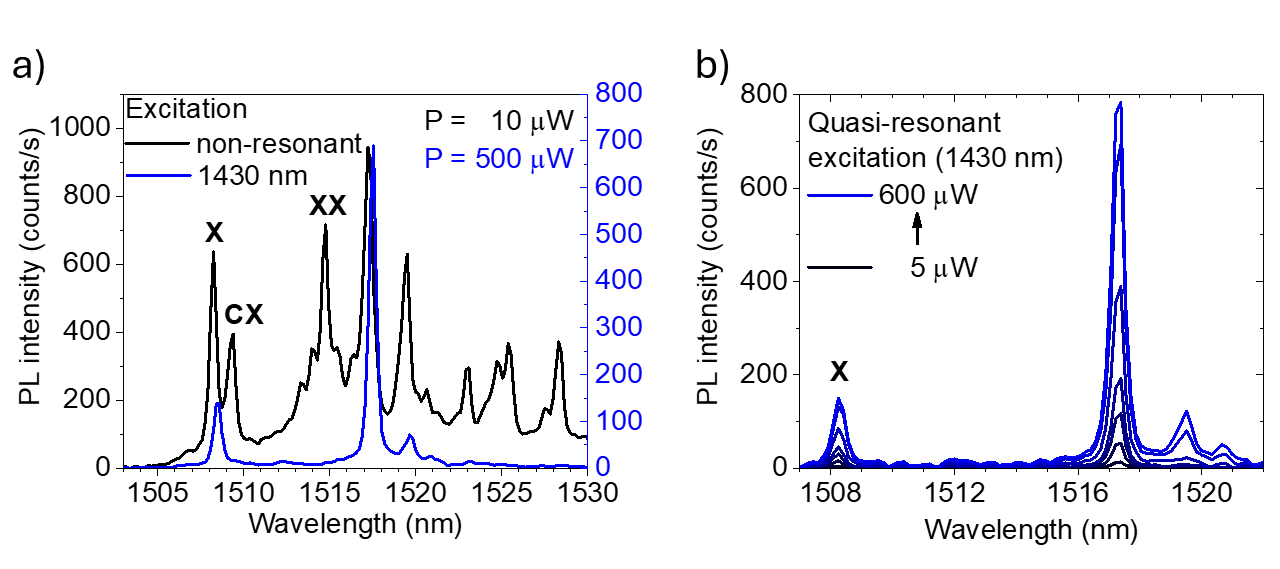}
  \caption{Low temperature (5K) {\textmu}PL spectra a) under non-resonant excitation with 10~{\textmu}W excitation power (black line) and quasi-resonant excitation at 1430 nm excitation wavelength with 500~{\textmu} (blue line); b) for various excitation powers in the range from 5~{\textmu}W to 600~{\textmu} (maximum available) at 1430 nm excitation wavelength}
  \label{fig:uPLE_spectra}
\end{figure}

Looking at the spectra under quasi-resonant excitation with 1430~nm wavelength shown in \textbf{Figure~\ref{fig:uPLE_spectra}}a, one can notice that the number of emission lines is limited, so there is no overlap between them that could add to the emission background, as typically observed. This opens up the possibility of creating higher-quality non-classical light sources when excited quasi-resonantly and still relatively far from the emission. Unfortunately, it is not possible to detect other emission lines from the QD under study, even at the highest available excitation power of 600~{\textmu}W (\textbf{Figure~\ref{fig:uPLE_spectra}b}). This shows that quasi-resonant excitation is not suitable for exciting the XX-X cascade in the case of limited excitation power and that neutral exciton is preferentially formed under this excitation scheme compared to the charged exciton. The simplest explanation for the latter is the lack of residual carriers in the QD before the excitation.
A charged exciton visible under non-resonant excitation, but not under quasi-resonant excitation, suggests that the trions are not formed due to the presence of a residual carrier in the QD, but due to different relaxation times for electrons and holes in the structure. This difference is due to the different electron and hole effective masses and can be enhanced by traps for one type of carrier in the surroundings of the QD. In a previous study, the presence of such traps was suggested by the increase in QD ensemble emission intensity as a function of temperature \cite{MusialPSS2023}.    

\subsection{Energy structure calculations}

\begin{figure}[tb]
  \includegraphics[width=0.55\linewidth]{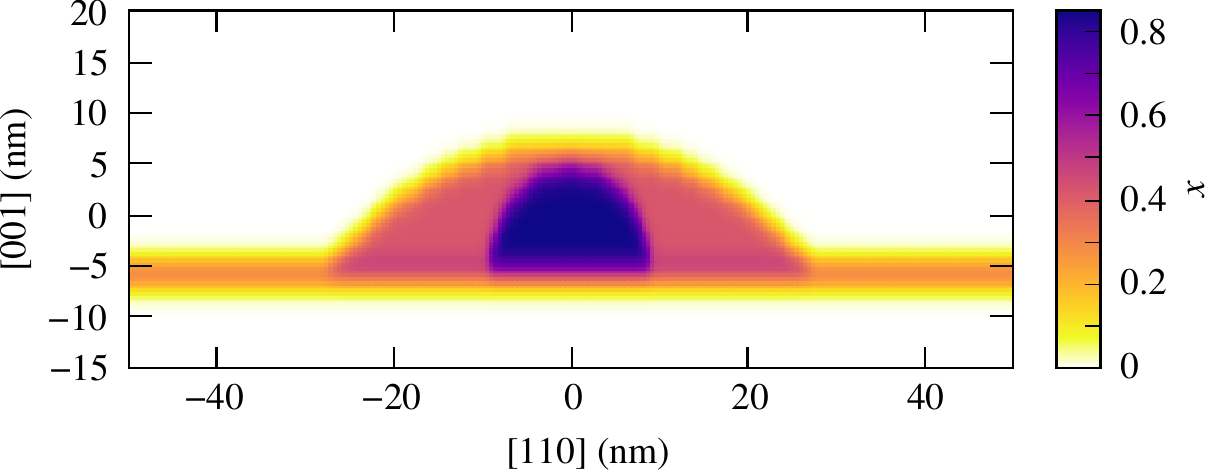}
  \caption{The model of the investigated QD. The color scale shows arsenic content $x$ in the InAs$_x$P$_{1-x}$ material.}
  \label{fig:QD-model}
\end{figure}

To support the interpretation of the experimental data, we calculate the energy structure of a typical QD of the investigated type. The modeling we conduct is similar to the one in Ref.~\cite{MusialPSS2023} and is based on the available structural data obtained from cross-sectional transmission electron microscopy (TEM) scans (\textbf{Figure~\ref{fig:TEM}b}), indicating that the QDs have a lens shape, with a base diameter of approximately 55~nm and a height of 10-15~nm (we assume 12~nm here) \cite{CarmesinPRB2017}. TEM images also show that QDs exhibit an inhomogeneous distribution of material, with a higher concentration of arsenic in the center. Based on this, we assume pure InP bulk material as the barrier, 85\% As concentration in the QD center, and 42\% in the outer part of the QD and in the 1.2-nm-thick wetting layer. \textbf{Figure~\ref{fig:QD-model}} shows a cross-section of the simulated material composition of the QD. We fix the used values by fine-tuning the ground-state energy to the observed emission within the uncertainty bounds for experimentally determined As concentrations. Then we apply Gaussian averaging (with a spatial extent of 0.9~nm) to simulate the interdiffusion of atoms at the interfaces.

We calculate the strain field within the continuum elasticity theory and the shear-strain-induced piezoelectric field. We then compute the eigenstates of electrons and holes using an implementation \cite{GawareckiPRB2014} of the $\bm{k}\cdot\bm{p}$ method \cite{BahderPRB1990}, which includes the effects of strain, piezoelectric field up to second-order terms in strain tensor elements, and spin-orbit coupling. The explicit form of the Hamiltonian used can be found in Ref.~\cite{Mielnik-PyszczorskiPRB2018}, and the material parameters in Refs.~\cite{HolewaPRB2020,GawelczykPRB2017} and references therein. To calculate the states of excitons and biexcitons, we use the configuration-interaction method \cite{BryantPRL1987}, with a basis of 48 electron and 48 hole eigenstates. Finally, we find the optical transition dipole moments, the resulting polarization of emission, and recombination times in the dipole approximation \cite{Andrzejewski_2010}.

\begin{figure}[tb]
  \includegraphics[width=0.5\linewidth]{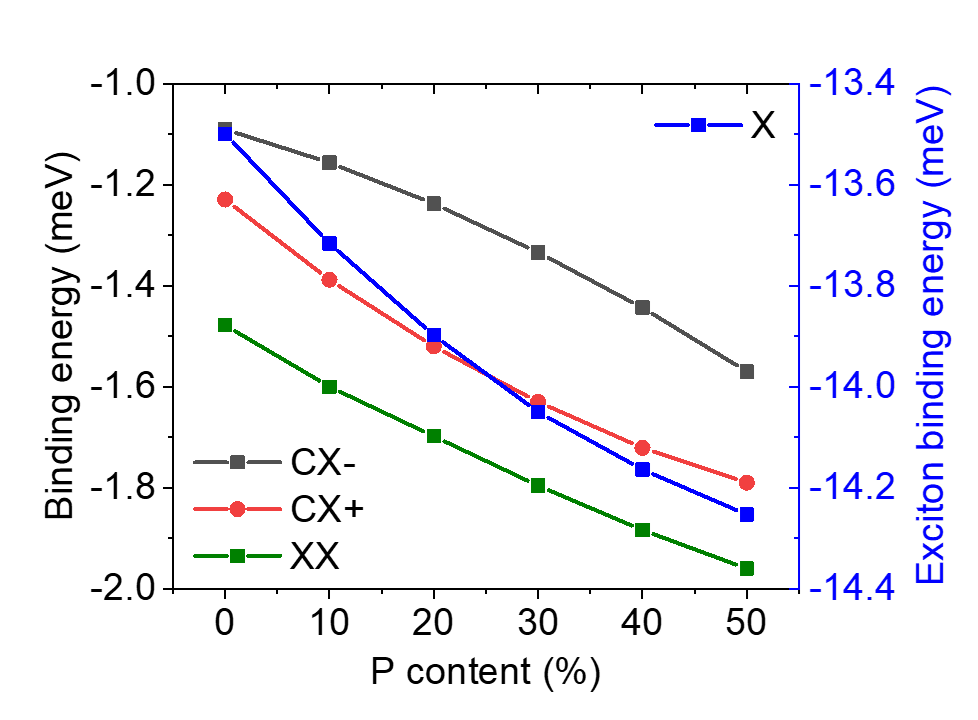}
  \caption{Calculated binding energies for exciton (blue, right-hand vertical axis), charged excitons (negatively - black and positively - red) and biexciton (green) - left-hand vertical axis, as a function of average phosphorus content within the QD volume}
  \label{fig:binding_energies}
\end{figure}

For such an exemplary QD, the energy difference between the two lowest-energy electron (heavy hole) orbital levels is $\sim50$~meV ($\sim19$~meV), which adds up to almost 70~meV energy distance from the ground state of the excitonic $p$-shell. This will be lowered by the exciton binding energy, which is $\sim14$~meV (\textbf{Figure~\ref{fig:binding_energies}}). This is qualitatively consistent with high-energy resonance in the \textmu{PLE} results and lack of further resonances when the energy distance is decreased. This large splitting is a result of highly inhomogeneous As composition within the QD volume (\textbf{Figure~\ref{fig:TEM}}), making the confining potential effectively narrower compared to what would be expected based on QD size. 

Other excitonic complexes may also form bound states (\textbf{Figure~\ref{fig:binding_energies}}), in agreement with the experimental findings, and the order of states suggests that the observed charged exciton is most probably of a negative character, as it features the lowest binding energy. Calculated binding energies increase with the phosphorus content, which would have to be increased beyond 50\% to match the experimentally obtained XX binding energy of 3.5~meV, but this leads to an increase in the emission energy beyond the range justifiable experimentally. It shows that the XX binding energy is underestimated, whereas charged excitons are bound more strongly than in the experiment, with similar trends as a function of QD material composition. One has to keep in mind that different exemplary QDs are studied via TEM and optical methods due to the destructive nature of TEM. So there is no direct correspondence between the structural and optical data (both are of exemplary character), and it should not be expected that the calculations and experimentally-determined parameters agree quantitatively. Comparing the emission energy of the QD under study (0.8212~eV) to the average emission energy from the QD ensemble (0.8052 eV), one immediately sees that this is not a typical QD, but can be analyzed owing to the fact that it comes from the tail of the QD distribution. We cannot evaluate how typical the TEM-based structural data are due to the limited statistics of these time- and resource-consuming measurements. 

\begin{figure}[tb]
  \includegraphics[width=\linewidth]{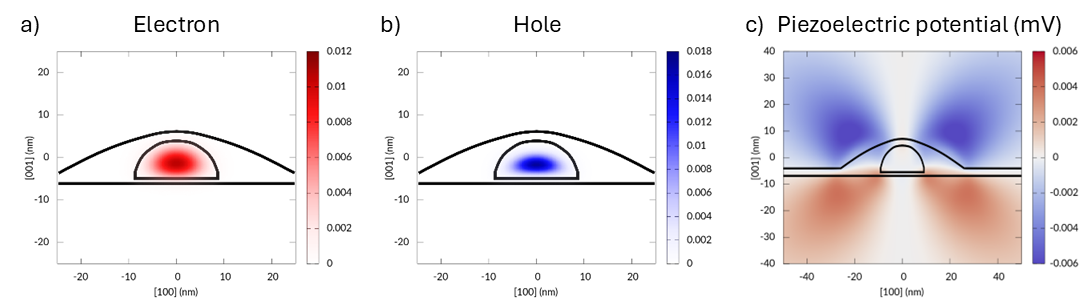}
  \caption{Calculated probability density distribution for electron (a) and hole (b) in two orthogonal planes; c) Calculated piezoelectric potential in the plane perpendicular to the growth plane with the contour of the QD shape and high As content region marked}
  \label{fig:piezo}
\end{figure}

In the current simulation, we treat the electron-hole exchange interaction parametrically (fine structure from experiment is fed into the calculations), so the FSS value from the previous section cannot be compared to a calculated one. However, one can understand the underlying mechanism for the residual non-zero FSS in a structurally symmetric QD. To investigate it, in \textbf{Figure~\ref{fig:piezo}} wave functions of the lowest electron and heavy hole states are presented together with the piezoelectric potential, which is typically the main reason for symmetry breaking in shape-symmetric strained QDs. This potential is localized mainly at the outer facets of the QD, while the wave functions are largely confined to the central region. For that reason, it is unlikely that piezoelectric potential alone is the main factor causing the observed FSS.
Our calculations do not account for alloy randomness, which can only be captured by atomistic approaches. It could explain the non-zero fine structure splitting observed in the experiment, as it is always present in ternary material alloys and can lead to single {\textmu}eV FSS \cite{Zielinski2020}.\\

\section{Entanglement}
\subsection{Quantum state tomography}

\begin{figure}[tb]
  \includegraphics[width=\linewidth]{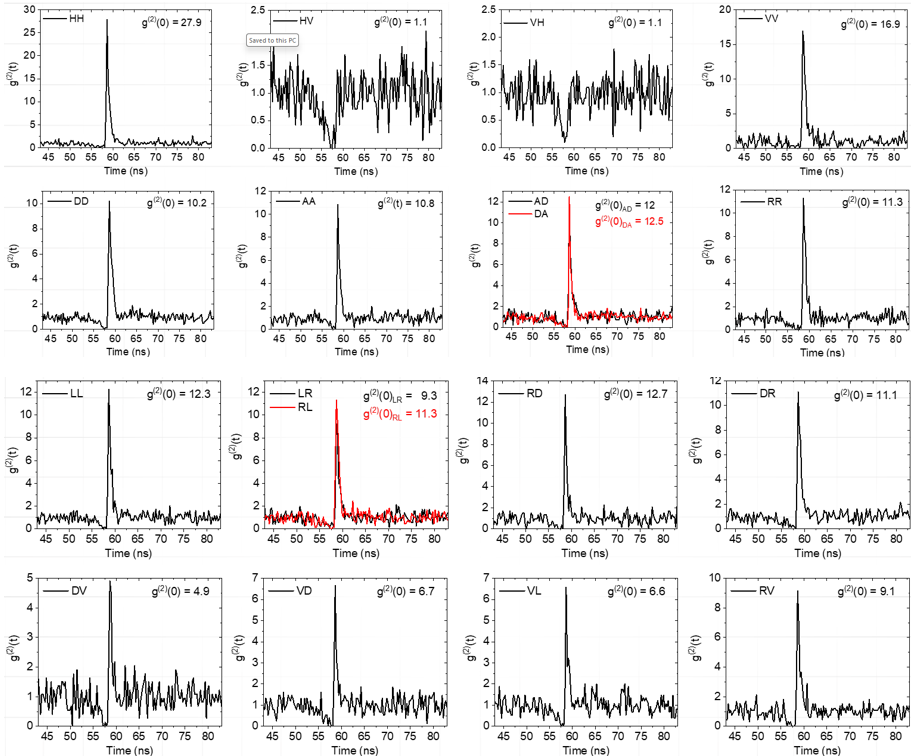}
  \caption{Quantum state tomography (experiment): second-order photon correlation measurements for the biexciton and exciton emission lines in different polarization bases indicated in each graph (the different polarization states are: V -- vertical, H -- horizontal, D -- diagonal, A -- anti-diagonal, R -- right-hand circular and L -- left-hand circular)}
  \label{fig:tomography}
\end{figure}

We perform quantum state tomography measurements \cite{james01} to determine the two-photon density matrix and estimate the entanglement fidelity of the pair of photons generated by the XX-X cascade. Measurements in four two-photon bases of states are carried out (HH, HV, VH, VV, DD, AA, DA, RR, LL, LR, DV, VD, RV, VL, DR, RD) with the first (second) letter indicating the polarization state onto which the photon resulting from the XX (X) recombination is projected (with the letters indicating the polarization: H -- horizontal, V -- vertical, D -- diagonal, and A -- antidiagonal are the different directions of linear polarization, and L -- left, R -- right refer to the circular polarization helicity). All 16 cross-correlations are measured, and the maximal value of $g^{(2)}(0)$ (bunching strength) from as-measured $g^{(2)}(\tau)$ data (without deconvolution with the resolution of the experimental setup) is taken to evaluate the two-photon density matrix. The strongest correlations are observed in the HV basis, with $g^{(2)}(0) = 27.9$ for the HH measurement and no bunching for the HV and VH measurements. The correlations are weaker, but still present in the DA and LR bases (\textbf{Figure~\ref{fig:tomography}}) as required for the two photons to be entangled. The obtained values are further used to determine the two-photon density matrix using the open-source software provided by Kwiat Quantum Information Group from the University of Illinois at Urbana-Champaign, available at: http://tomography.web.engr.illinois.edu/TomographyDemo.php. The reconstructed two-photon density matrix provides input for analyzing the entangled state described in the following Subsection. 

\subsection{Analysis of the measured entangled state}

The state under study, written in density matrix form, is given by
\begin{equation}
	\label{stan}
	\hat{\rho}=\left(
	\begin{array}{cccc}
		a&x&y&z\\
		x^*&b&p&q\\
		y^*&p^*&c&s\\
		z^*&q^*&s^*&d
	\end{array}	
	\right),
\end{equation}
with the diagonal elements given by $a=0.6$, $b=0.238$, $c=0.0212$, and $d=0.354$,
while the off-diagonal elements are $x=-3.16\times 10^{-3}+0.0789 i$, $y=0.0815-0.0267 i$, $z=-0.0237-0.0514 i$, $p=6.18\times 10^{-4}-0.0144 i$, $q=-0.0731-5.56\times 10^{-3} i$, and $s=-0.0302-0.0302 i$.
The above matrix is given in the two-qubit computational basis, 
in the standard order, $\{\lvert00\rangle$, $\lvert01\rangle$, $\lvert10\rangle$, $\lvert11\rangle\}$.
The concurrence \cite{wootters98} of this state is $C(\hat{\rho})=0.145$, so it is entangled 
(keeping in mind that $C(\hat{\rho})\in[0,1]$, with $C(\hat{\rho})=0$
signifying a separable state).

To understand the nature of the entanglement present in the state, it is helpful to find the pure state decomposition of state $\hat{\rho}$, which minimizes 
the average of entanglement present in the pure states.
To this end, it is more straightforward to work with Entanglement of Formation (EoF)
due to its definition:
EoF of a mixed state $\hat{\rho}$ is given by the average
entanglement entropy of the pure states that compose the state minimized
over all possible pure state decompositions of $\hat{\rho}$ \cite{bennett96,bennett96a}.
A similar definition involving a convex roof construction applies to the two-qubit concurrence, but it requires the use of 
tangle as the pure state qubit measure \cite{hildebrand07}.
For two qubits, there is a direct formula to transform concurrence into EoF \cite{wootters98},
\begin{equation}
	E(\hat{\rho})=-x_+\log_2(x_+)-x_-\log_2(x_-),
\end{equation}
with 
\begin{equation}
	x_{\pm}=\frac{1\pm\sqrt{1-C^2(\hat{\rho})}}{2},
\end{equation}
which yields $E(\hat{\rho})=0.0476$ for state (\ref{stan})
and the more commonly used pure state entanglement measure can be used. 

We start by diagonalizing the state (\ref{stan}) and obtain the set of 
eigenvalues $p_i$, with $i=0,1,2,3$ and corresponding eigenstates
\begin{equation}
	\lvert\psi_i\rangle=\alpha_i\lvert00\rangle+\beta_i\lvert01\rangle+\gamma_i\lvert10\rangle+\delta_i\lvert11\rangle.
\end{equation}
The eigenvalues are given by $p_0=0.6402$, $p_1=0.3535$, $p_2=0.0063$, and $p_3=0$, and the 
coefficients of the corresponding eigenstates are given in Table~\ref{coef}.
The decomposition of $\hat{\rho}$ into its eigenstates is not the decomposition which 
minimizes average entanglement since 
\begin{equation}
	\label{av}
	\sum_{i=0}^3p_iE(\lvert\psi_i\rangle)=0.3490
\end{equation}
vastly exceeds the EoF of the state. Here $E(\lvert\psi_i\rangle)$ denotes the entanglement entropy
of the pure state and it is equal to $0.3544$, $0.3376$, $0.4328$, $0.4242$,
respectively for $i=0$, 1, 2, 3.

\begin{table}
	\caption{\label{coef} Table of coefficients for the eigenstates of the density matrix (\ref{stan}).}
	\begin{tabular}[tbp]{@{}|c|c|c|c|c|@{}}
		\hline		
		$i$&$\alpha_i$&$\beta_i$&$\gamma_i$&$\delta_i$\\
		\hline
		0&-0.9475&-0.0096+0.1498 i&-0.1437-0.0349 i	&0.0969-0.2201 i\\
		1&0.2669&-0.0971+0.1254i&-0.0531+0.0628i&0.3720-0.8709i\\
		2&0.1527&0.2309+0.1903i&-0.7972-0.4975i&-0.0503+0.0397i\\
		3&0.0878&-0.2096+0.9049i&-0.0033+0.2971i&0.0018+0.2030i\\
		\hline
	\end{tabular}
\end{table}

Note that one of the eigenvalues is zero, while the other is very small in comparison with
$p_0$ and $p_1$. Hence, in the following analysis, we will approximate the state $\hat{\rho}$
by
\begin{equation}
	\label{tilde}
	\tilde{\rho}=\tilde{p}_0\lvert\psi_0\rangle\langle\psi_0\rvert+\tilde{p}_1\lvert\psi_1\rangle\langle\psi_1\rvert,
\end{equation}
where the probabilities are adjusted to $\tilde{p}_i=p_i/N$ and $N=p_0+p_1$,
yielding $\tilde{p}_0=0.6443$ and $\tilde{p}_1=0.3557$.
The limitation of the effective Hilbert space of the state under study will significantly simplify the minimization required to find the pure state decomposition yielding
the amount of entanglement in the system. On the other hand, the error made is minimal 
as the fidelity between the two states is 
\begin{equation}
	F(\hat{\rho},\tilde{\rho})=\left(\mathrm{Tr}\sqrt{\sqrt{\hat{\rho}}\tilde{\rho}\sqrt{\hat{\rho}}}\right)^2=0.9937.
\end{equation}
Since the fidelity is equal to one for identical states and to zero for orthogonal states
(or, to be precise, when dealing with mixed states, states that have orthogonal supports),
we see that the difference between the two states is negligible, since the fidelity is over
$99 \%$.
The average entanglement of the decomposition of this state into its 
eigenstates is $\sum_{i=0}^1\tilde{p}_iE(|\psi_i\rangle)=0.3484$, meaning that it is almost
identical to the average entanglement of state (\ref{stan}), given by Eq.~(\ref{av}).

We would now like to find the pure state decomposition that minimizes average pure state entanglement
for the state (\ref{tilde}). We will restrict our consideration to the two-state subspace of 
$\{\lvert\psi_0\rangle,\lvert\psi_1\rangle\}$, since this drastically reduces the complexity of
the problem. The price to pay is that we will not find the exact state that minimizes
entanglement, since the states $\lvert\psi_0\rangle$ and $\lvert\psi_1\rangle$ are not
confined to a separable two-state subspace.
This can be seen from the respective Schmidt decompositions of the two states,
\begin{equation}
	\label{schmidt}
	\lvert\psi_i\rangle =\sum_{j=0}^1\sqrt{p_j^{(i)}}e^{i\varphi_j^{(i)}}|\psi_j^{(i)}\rangle_A
	\otimes\lvert\psi_j^{(i)}\rangle_B.
\end{equation}
For state $\lvert\psi_0\rangle$, the parameters of the decomposition are
$p_0^{(0)}=0.9331$ with corresponding states 
	\begin{eqnarray}
		\label{s1a}
\lvert\psi_0^{(0)}\rangle_A&=&0.9926 e^{-i 0.0635\pi}\lvert0\rangle_A-0.1215\lvert1\rangle_A,\\
\label{s1b}
|\psi_0^{(0)}\rangle_B&=&0.9922 e^{i 0.4915\pi}\lvert0\rangle_B-0.1246\lvert1\rangle_B,
\end{eqnarray}
and $p_1^{(0)}=0.0669$ where the corresponding states are orthogonal to $\lvert\psi_0^{(0)}\rangle_A$
and $\lvert\psi_0^{(0)}\rangle_B$, respectively.
For state $\lvert\psi_1\rangle$, the parameters are
$p_0^{(1)}=0.9374$ with corresponding states 
\begin{eqnarray}
	\label{s2a}
	\lvert\psi_0^{(1)}\rangle_A&=&0.1966 e^{i 0.0626\pi}\lvert0\rangle_A-0.9805\lvert1\rangle_A,\\
	\label{s2b}
	\lvert\psi_0^{(1)}\rangle_B&=&0.1327 e^{i 0.0423\pi}\lvert0\rangle_B-0.9912\lvert1\rangle_B,
\end{eqnarray}
and $p_1^{(1)}=0.0133$ with orthogonal corresponding states, as before.
   
We follow the 
prescription found in Ref.~\cite{nielsen00} to find all possible pure-state decompositions
within the confinement of the two states $\lvert\psi_0\rangle$ and $\lvert\psi_1\rangle$. 
Those are given by
\begin{equation}
	\tilde{\rho}=q_0\lvert\phi_0\rangle\langle\phi_0\rvert+q_1\lvert\phi_1\rangle\langle\phi_1\rvert,
\end{equation}
with
\begin{eqnarray}
	\label{deca}
	\lvert\phi_0\rangle&=&\frac{1}{q_0}\left[\alpha\sqrt{\tilde{p}_0}\lvert\psi_0\rangle
	+\beta\sqrt{\tilde{p}_1}\lvert\psi_1\rangle\right],\\
	\label{decb}
	\lvert\phi_1\rangle&=&\frac{1}{q_1}\left[\beta^*\sqrt{\tilde{p}_0}\lvert\psi_0\rangle
	-\alpha^*\sqrt{\tilde{p}_1}\lvert\psi_1\rangle\right],
\end{eqnarray}
and $q_0=\lvert\alpha\rvert^2\tilde{p}_0+\lvert\beta\rvert^2\tilde{p}_1$, $q_1=1-q_0$,
for any parameters $\alpha$ and $\beta$ that fulfill $\lvert\alpha\rvert^2+\lvert\beta\rvert^2=1$.

The parameters that minimize entanglement are $\alpha=-0.2480$ and $\beta=0.9688 e^{i 0.9954\pi}$,
yielding $q_0=0.3735$ and $q_1=0.6265$
and the respective values of the EoF, $E(\lvert\phi_0\rangle)=0.0497$ and $E(\lvert\phi_1\rangle)=0.0507$.
This means that the average value of entanglement for this decomposition of the state is 
$\sum_{i=0}^1q_iE(\lvert\phi_i\rangle)=0.0503$, overestimating the value of the EoF of state (\ref{stan})
by only $5.7\%$, as opposed to the average EoF in the eigendecomposition, which overestimates by $633\%$
(is $7.33$ times larger). This suggests that the decomposition found is a reasonable approximation
of the decomposition which truly minimizes entanglement, and we will finally give the Schmidt decomposition
of the states (\ref{deca}) and (\ref{decb}) with the specified $\alpha$ and $\beta$ to understand entanglement present
in those states.
This is given by a construction analogous to that given by eq.~(\ref{schmidt}) and for the state (\ref{deca})
we get 
$p_0^{(0)}=0.9944$ with corresponding single-qubit states 
	\begin{eqnarray}
		\label{d1a}
		\lvert\psi_0^{(0)}\rangle_A&=&0.8735 e^{-i 0.9682\pi}\lvert0\rangle_A-0.9823\lvert1\rangle_A,\\
		\label{d1b}
		\lvert\psi_0^{(0)}\rangle_B&=&0.0966 e^{-i 0.7887\pi}\lvert0\rangle_B-0.9953\lvert1\rangle_B,
	\end{eqnarray}
and $p_1^{(0)}=0.0056$ corresponding to states orthogonal to the ones above. 
Analogously for state (\ref{decb}), the parameters are
$p_0^{(1)}=0.9943$ with corresponding states 
	\begin{eqnarray}
		\label{d2a}
		\lvert\psi_0^{(1)}\rangle_A&=&0.9913 e^{i 0.1038\pi}\lvert0\rangle_A-0.1319\lvert1\rangle_A,\\
		\label{d2b}
		\lvert\psi_0^{(1)}\rangle_B&=&0.9933 e^{i 0.5264\pi}\lvert0\rangle_B-0.1151\lvert1\rangle_B,
	\end{eqnarray}
and $p_1^{(1)}=0.0057$ with orthogonal corresponding states, as before.

Once the Schmidt decomposition is found, quantification of entanglement only depends on the probabilities 
$p_j^{(i)}$ as both single qubit bases in Eq.~(\ref{schmidt}) can be changed to any other single qubit
basis (including the computational basis $\{\lvert0\rangle,\lvert1\rangle\}$) via local unitary operations.
This is also the case for the phases present in Eq.~(\ref{schmidt}).
Entanglement entropy for state $\lvert\phi_i\rangle$ can, in fact, be written as 
\begin{equation}
	E(\lvert\phi_i\rangle)=-\sum_{j=0}^1p_j^{(i)}\log_2p_j^{(i)},
\end{equation}
and this form makes it evident that the value of entanglement is greatest for $p_0^{(i)}=p_1^{(i)}=1/2$,
whereas the closer either of the probabilities is to zero, the smaller the entanglement in the pure state.
This explains why the states obtained via the decomposition (\ref{deca},\ref{decb}) have so little entanglement 
in comparison with the eigenstates of $\tilde{\rho}$, since the smaller probabilities $p_1^{(i)}$
are an order of magnitude closer to zero for decompositions (\ref{d1a},\ref{d1b}) and (\ref{d2a},\ref{d2b}) - $p_1^{(0)}=0.0056$, $p_1^{(1)}=0.0057$ - 
than the Schmidt decompositions of the eigenstates (\ref{s1a},\ref{s1b}) and (\ref{s2a},\ref{s2b}) - $p_1^{(0)}=0.0669$, $p_1^{(1)}=0.0133$.
 
To summarize the calculation, we have approximated the experimentally obtained entangled state (\ref{stan}) by a state
that incorporates two of the four eigenstates of (\ref{stan}), given by eq.~(\ref{tilde}). The fidelity between the two
states is over $99\%$, so it is a good approximation. This allowed us to find the pure state decomposition that yields EoF directly from the definition, obtaining $E(\tilde{\rho})=0.0503$, which is comparable to the true entanglement of the
measured state, $E(\hat{\rho})=0.0476$. The decomposition consists of a mixture of two non-orthogonal states that are
much less entangled than the two dominant eigenstates of the density matrix, as seen in the respective Schmidt 
decompositions of both pairs of states.

\section{Conclusion}
In this work, we focus on neutral excitonic complexes confined in in-plane symmetric InAs(P)/InP quantum dots emitting in the third telecommunication window. We provide a comprehensive experimental and theoretical description of neutral excitonic complexes in these application-relevant structures. Their confinement strength is weaker than that of small QDs emitting at shorter wavelengths. This has consequences for the carrier dynamics (X to XX lifetime ratio of 1.5) and the importance of excitonic effects in these QDs (binding energies of 14 meV and 3.5 meV, for exciton and biexciton, respectively). The residual fine structure splitting is most likely related to the alloy randomness inevitable in ternary material QDs and not the piezoelectric potential, as carriers are confined to a central QD region, almost not overlapping with the piezoelectric field. The experimentally observed fine structure splitting $< 10$~{\textmu}eV allowed for the generation of entangled photon pairs from the biexciton-exciton recombination cascade. 

We found that the amount of entanglement in the photon pairs, quantified by Entanglement of Formation,
is much smaller than expected on the basis of the overall coherence of the state, as well as its eigendecomposition.
To understand this, we have minimized entanglement over all possible pure state decompositions in a limited two-dimensional
subspace. The fidelity of the studied state is over $99\%$ with respect to the measured state, and the entanglement
of the optimal decomposition exceeded the Entanglement of Formation of that state by only $5.7\%$, meaning that 
the two states are extremely similar. The two pure non-orthogonal states that enter the decomposition minimizing entanglement are much less entangled than the states in the eigendecomposition. This means that the decoherence and phase accumulation in the experiment interplay in an unfortunate way that reduces entanglement.

\medskip
\textbf{Acknowledgements} \par 
The research was financially supported by the QuanterERA II European Union’s Horizon 2020 research and innovation programme under the EQUAISE project, Grant Agreement No. 101017733, under the supervision of the National Center for Research and Development in Poland within the project QuantERA II ERA-Net Cofund in Quantum Technologies (QUANTERAII/1/74/EQUAISE/2022)
and under the QuCABOoSE project, Grant Agreement No. 101017733, under the supervision of
MEYS (Czechia), as well as Bundesministerium für Forschung, Technologie und Raumfahrt-BMFTR in the frame of the project QR.N (16KIS2204) and DFG-Heisenberg grant (BE 5778/4-1). 
KR is funded within the QuantERA II Programme that has received funding
from the EU H2020 research and innovation programme
under Grant Agreement No. 101017733, and with
funding organization MEYS.
Calculations have been carried out using resources provided by Wroc{\l}aw Centre for Networking and Supercomputing (http://wcss.pl), Grant No. 203. We are grateful to Krzysztof Gawarecki for sharing his $\bm{k}\cdot\bm{p}$ method implementation. We also acknowledge Andrei Kors for his assistance in the MBE growth process, Kerstin Fuchs, and Dirk Albert for technical support.


\medskip

%
\bibliographystyle{MSP}
\bibliography{manuscript}




\end{document}